\documentclass[a4paper]{article}
\usepackage[style=nature]{biblatex}
\usepackage[affil-it]{authblk}
\usepackage{amsmath}
\usepackage{amsthm}
\usepackage{amssymb}
\usepackage[margin=1in]{geometry}
\usepackage[utf8]{inputenc}
\usepackage{hyperref}
\usepackage{subcaption}
\usepackage{graphicx, color}
\usepackage[english]{babel}
\usepackage{blindtext}
\usepackage{algorithm, algpseudocode} 
\usepackage{mathrsfs} 
\usepackage{hhline}
\usepackage{setspace}
\hypersetup{
    colorlinks=true,
    allcolors=blue,
    }
\renewcommand{\thefootnote}{\fnsymbol{footnote}}

\title{\textbf{Superior mechanical properties by exploiting size-effects and multiscale interactions in hierarchically architected foams}}

\author[1]{Komal Chawla$^{\dagger,}$}
\author[2]{Abhishek Gupta$^{\dagger,}$}
\author[3]{Abhijeet S. Bhardwaj}
\author[1,2]{Ramathasan Thevamaran $^{*,}$}

\affil[1]{Department of Engineering Physics, University of Wisconsin-Madison, Madison, WI, 53706, USA}
\affil[2]{Department of Mechanical Engineering, University of Wisconsin-Madison, Madison, WI, 53706, USA}
\affil[3]{Department of Industrial and Systems Engineering, University of Wisconsin-Madison, Madison, WI, 53706, USA}

\date{\today}

\begin{document}
	\maketitle
	\sloppy
	\onehalfspacing
	\begin{abstract}

Protective applications in extreme environments demand thermally stable materials with superior modulus, strength, and specific energy absorption (SEA) at lightweight. However, these properties typically have a trade-off. Hierarchically architected materials---such as the architected vertically aligned carbon nanotube (VACNT) foams---offer the potential to overcome these trade-offs to achieve synergistic enhancement in mechanical properties. Here, we adopt a full-factorial design of experiments (DOE) approach to optimize multitier design parameters to achieve synergistic enhancement in SEA, strength, and modulus at lightweight in VACNT foams with mesoscale cylindrical architecture. We exploit the size effects from geometrically-confined synthesis and the highly interactive morphology of CNTs to enable higher-order design parameter interactions that intriguingly break the diameter-to-thickness (D/t)-dependent scaling laws found in common tubular architected materials. We show that exploiting complementary hierarchical mechanisms in architected material design can lead to unprecedented synergistic enhancement of mechanical properties and performance desirable for extreme protective applications.
		

\end{abstract}

\def\thefootnote{$\dagger$}\footnotetext{These authors contributed equally to this work}
\def\thefootnote{$*$}\footnotetext{Corresponding author, Email address: thevamaran@wisc.edu}

	\doublespacing
	
	\section{Introduction}

Exceptional modulus, strength, and toughness with lesser mass density and thermal stability in extreme conditions are critical characteristics of materials that are required in aerospace, automotive, robotics, structural, and defense applications. These qualities, however, are not found together in a conventional material due to the contradictory relationship they have with material's morphology and the different lengthscales of their origin \autocite{lakes1993materials,ritchie2011conflicts}. For example, stiffer materials are typically poor in dissipating energy \autocite{lakes2009viscoelastic}, and high-strength materials generally have lesser toughness \autocite{ritchie2011conflicts, fang2011revealing,thevamaran2016dynamic}.  Materials with structural hierarchy have been shown to overcome the limitations on mechanical performance by exploiting superior intrinsic properties of the nanoscale building blocks and their interactions across the structural organization from nano to macroscales to achieve high strength, toughness, and Young's modulus at lightweight than their constituent materials \autocite{lakes1993materials,ortiz2008bioinspired,munch2008tough,tang2003nanostructured,chen2012biological,meza2015resilient,yin2019impact}. Multiscale hierarchy and functional property gradients are found in biological materials and have been a source of inspiration to develop synthetic hierarchical materials with functionalities tuned for specific needs \autocite{ortiz2008bioinspired,munch2008tough,tang2003nanostructured,chen2012biological,yin2019impact,lee2014hierarchical,jia2019biomimetic,miserez2008transition,cao2005super,lin2015biomimetic,thevamaran2015shock}. Similarly, architected materials with specific geometries---sometimes referred to as mechanical metamaterials---have also been utilized to achieve superior specific properties and unusual functionalities, for example, multistable structures for energy dissipation \autocite{shan2015multistable,haghpanah2017elastic}, nacre like structure for improved fracture toughness \autocite{ritchie2011conflicts,yin2019impact}, nanotruss lattices for improved specific stiffness \autocite{meza2017reexamining,schaedler2011ultralight}, cellular and lattice structures for improved thermal properties \autocite{lakes1996cellular,yamamoto2014thin} and recently PT symmetric fractals for anomalous wave transport \autocite{fang2021universal}. These studies suggest that exploiting the interplay between multitier geometric parameters of an architected design along with structural hierarchy present in a material system like vertically aligned carbon nanotube (VACNT) foams can yield unprecedented bulk properties desirable for protective applications. Additionally, they can serve as versatile design templates for hierarchical materials to achieve synergistic property enhancement and desired functionalities.

Here, we report synergistic improvement in specific elastic modulus, specific energy absorption, and specific compressive strength in architected VACNT foams. The VACNT foams have a hierarchical structure with feature sizes ranging from a few angstroms to several millimeters \autocite{cao2005super}. The individual CNT's multiwalled structure at the nanoscale along with their entangle forest-like morphology in microscale that further forms into vertically aligned bundles in mesoscale make them undergo collective sequencially progressive buckling under compressive loading and exhibit bulk strain recovery of over 80\% \autocite{cao2005super, gupta2022origins}. They also exhibit superior thermal stability of their mechanical properties from -196 to 1000 $^{\circ}C\;$ \autocite{xu2010carbon}. The hierarchical structure of VACNT foams can be tailored to affect their bulk properties dramatically by varying synthesis parameters \autocite{raney2011tailoring,yaglioglu2012wide,thevamaran2015shock,thevamaran2015anomalous,gupta2022origins} and by introducing micropatterned growth process \autocite{de2010diverse,copic2011fabrication,lattanzi2014geometry,lattanzi2015dynamic,jeong2009effect}. We introduced an additional level of structural hierarchy in our VACNT samples through an architected hexagonally packed lattice of hollow cylinders in mesoscale (in the order of 100 $\mu m$) (\autoref{fig:her1}). Hollow cylinders provide additional structural rigidity to the samples \autocite{farley1986effect,alia2015energy}, and the hexagonal pattern allows close packing that enhances interactions among the neighboring cylinders. We design, synthesize, and analyze multiple samples by varying and intermixing the three geometrical parameters---the inner diameter of cylinders $(D_{in})$, the thickness of cylinders $(t)$, and the gap between the outer walls of neighboring cylinders $(g)$.

To achieve simultaneous improvement in mechanical properties, we used a full factorial design of experiments (DoE) approach with $D_{in}$, $t$, and $g$ as our design variables. In contrast to the one variable at a time (OVAT) approach, where changing one variable leave the effect of other variables and their interconnectivity unforeseen, full factorial design allows more comprehensive multivariable study and reveals the correlation between design variables \autocite{cao2018optimize}. We choose specific energy absorption ($SEA$), specific modulus ($E^*$), and specific peak stress ($\sigma_p^*$) calculated from quasistatic stress-strain curve as our response variables (objectives). We observe synergistic improvement in $SEA$, $E^*$, and $\sigma_p^*$ owing to morphology changes due to geometrically-confined CNTs growth (size effects), lateral interactions among adjacent cylinders, and the relationship among design variables revealed by fitted Analysis of Variance (ANOVA) models. Our material transcends the traditional requirements for protective applications and demonstrates a design template for architected materials to achieve desired properties

\section{Architected Design}

\begin{figure}[t]
	\centering
	\includegraphics[width=1\textwidth]{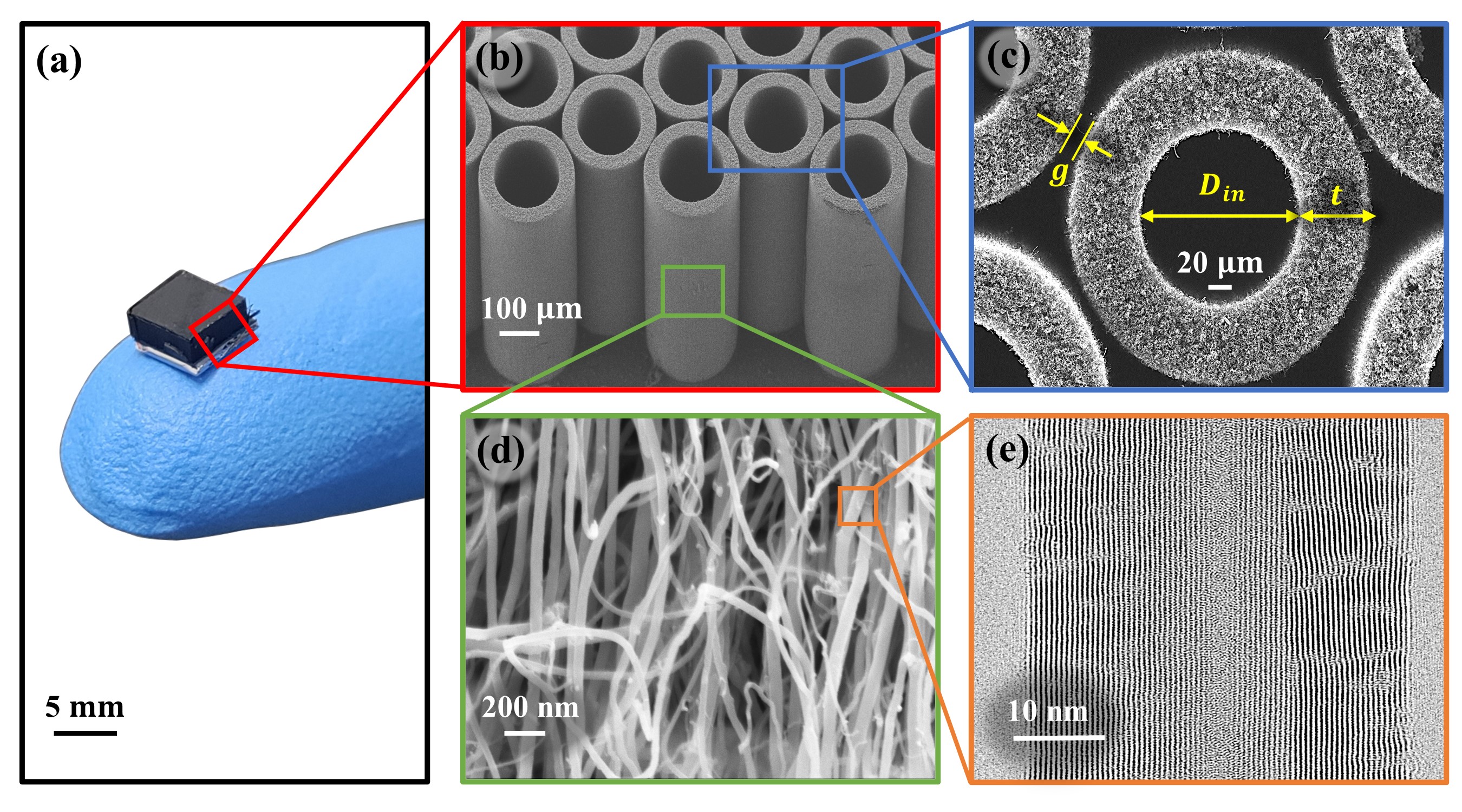}	\caption{(a) CVD synthesized pristine VACNT sample on a $5\;mm \times 5\;mm$ substrate. (b,c) SEM images showing the mesoscale hexagonally-packed hollow cylinders made of vertically aligned carbon nanotubes; inner diameter ($D_{in}$) and thickness ($t$) of a cylinder and the gap ($g$) between adjacent cylinders are indicated. (d) An SEM image showing the entangled forest-like morphology of vertically-aligned individual CNTs in microscale. (e) A TEM image showing the multi-walled structure of an individual CNT at the nanoscale.}
	\label{fig:her1}
\end{figure}

We adopt a full factorial design of experiments (DoE) approach to optimize the mechanical performance of microarchitected VACNT foams as a function of the geometric parameters. \autoref{fig:her1} shows electron microscopy images of a VACNT foam sample with hexagonal close-packed cylindrical architecture synthesized in a floating-catalyst chemical vapor deposition (CVD) process. The multilevel hierarchy is apparent from the continuum-like appearance of structural features at each subsequently larger length scale---a distinction that separates hierarchical materials from conventional materials. To synthesize these samples, we use a floating catalyst CVD method to grow nearly vertically aligned bundles of MWCNTs (\autoref{fig:her1}(d,e)) on a photolithographically pre-patterned substrate (See \hyperref[section:sd2]{Methods}). Synthesized samples reach bulk height of $\sim 1.6 \;mm$ (\autoref{fig:her1}(a)) with a mesoscale cylindrical architecture of cross-sectional dimensions defined by the pre-patterned substrate. 

\begin{figure}[t]
	\centering
	\includegraphics[width=0.9\textwidth]{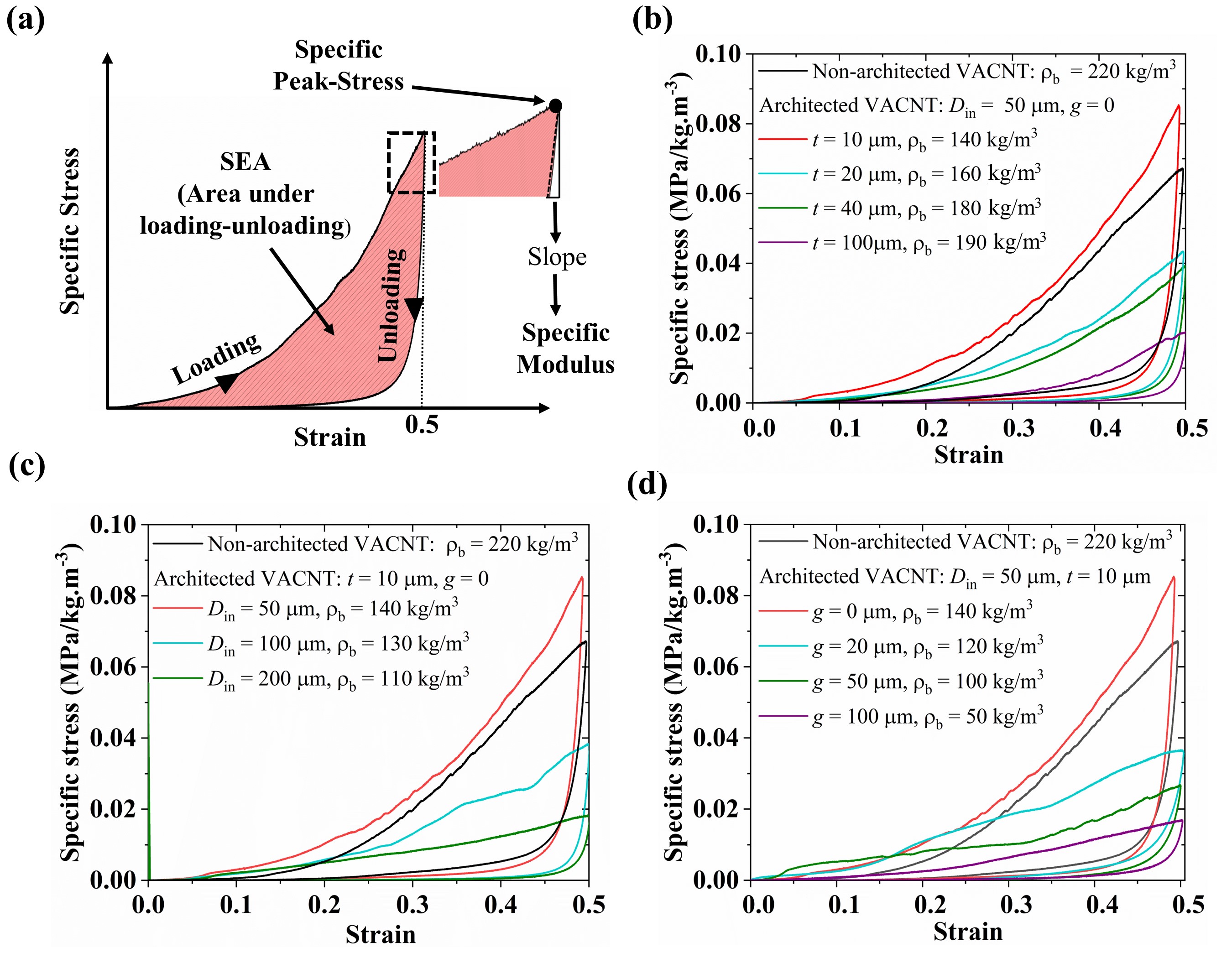} \caption{(a) Specific stress-strain response of VACNT and mechanical properties---SEA, $\sigma_p^*$, and $E^*$. (b) Specific stress-strain response as a function of $t$ for $D_{in}=50\;\mu m$ and $g=0$ compared to that of bulk (non-architected) VACNT sample. (c) Specific stress-strain response as a function of $D_{in}$ for $t=10\;\mu m$ and $g=0$. (d) Specific stress-strain response as a function of $g$ for $D_{in}=50\;\mu m$ and $t=10\; \mu m$. Stress-strain responses without density normalization are given in \hyperref[section:sd]{SI} Figure S1}
	\label{her2}
\end{figure}

The design variables we adopted for our DoE study---internal diameter $(D_{in})$, thickness $(t)$, and gap between cylinders $(g)$ are shown in \autoref{fig:her1}(c). We synthesized 180 samples with three in each of the 60 different combinations of $D_{in}$, $t$, and $g$ listed in \autoref{tab1}. The three tested samples in each unique combination of design variables allow us to characterize any variability in the microstructure of CNTs within cylinders resulting from the CVD synthesis process and mechanical testing. These variations are represented by error bars in the data presented in figures. Multiple levels (\autoref{tab1}) considered for each design variable allow us to study both their individual and combined effects on the stress-strain responses and associated mechanical properties. We performed cyclic quasi-static (strain rate: $0.01\;sec^{-1}$) compressions of up to $50\%$ strain on all samples using an Instron Electropulse E3000 apparatus to characterize the stress-strain behavior. \autoref{her2}(a) shows a typical stress-strain response with energy dissipated given by the hysteresis area enclosed by the loading-unloading curves. Protective applications require high specific properties, i.e., higher properties at low densities, so we normalized stress by the bulk density ($\rho_b$) of the sample to measure the specific stress response with compressive strain (\autoref{her2}). We pursue three functional objectives (or response variables) to simultaneously maximize---specific energy absorption ($SEA$), specific peak stress ($\sigma_p^*$), and specific modulus ($E^*$)---which are measured from the specific stress-strain curve (\autoref{her2}). We measure the SEA from the specific stress-strain hysteresis corresponding to the cyclic compression up to $50\%$ strain, which is much less than the typical densification strain ($\sim65-70\%$) in VACNTs \autocite{cao2005super,raney2011tailoring,thevamaran2015shock}. $\sigma_p^*$ indicates the specific compressive stress corresponding to the $50\%$ strain. It should be noted that the compressive strength---the stress at which the material permanently fails---of the VACNT foams is much higher as they have the ability to withstand and near completely recover from $80\%$ or higher strains. $E^*$ denotes the specific modulus typically measured from the elastic unloading of the foam-like materials. We evaluate $E^*$ by calculating the average slope of the unloading curve for the first $30\%$ of unloading in stress.

\begin{table}[H]
\centering
\caption{Design variables and their different levels.}
\resizebox{0.45\textwidth}{!}{%
\begin{tabular}{cccc}
\hhline{====}
\textbf{Design Variables} & \textbf{Levels} \\ \hhline{====}
$\mathbf{D_{in} (\mu m)}$               & $50, 100, 200$             \\ \hline
$\mathbf{t (\mu m)}$               & $10, 20, 40, 100$             \\ \hline
$\mathbf{g (\mu m)}$               & $0, 20, 50, 100, 200$              \\ \hline
\end{tabular}%
}\label{tab1}
\end{table}

\section{Results and discussion}

When compressed, VACNT bundles start to buckle in the bottom region (less dense and less stiff region close to the substrate) and then the buckles sequentially propagate upwards \autocite{cao2005super,hutchens2010situ}, causing local densification and nonlinear stiffening---a response typically described by a series of bi-stable elements consisting of an unstable phase in between \autocite{fraternali2011multiscale,thevamaran2014multiscale} or by hardening-softening-hardening plasticity \autocite{needleman2012deformation,liang2017compression}. This progressive sequential buckling response in VACNTs synthesized by floating-catalyst CVD is strongly governed by the mass density gradient across the height of the samples resulting from synthesis---entangled morphology evolution and increasing CNT population by continuous nucleation of new CNTs as the sample grows lead to the local mass density to increase from the substrate towards the top of the sample. Upon unloading, the sample recovers almost completely, exhibiting an enormous amount of hysteretic energy dissipation---$\sim83\%$ of total energy, corresponding to a damping capacity of $83\%$. In our architected VACNT samples, the damping capacity of $\sim83\%$ persists for all combinations of $D_{in}$, $t$, and $g$, which suggest that the bulk mechanical performance of VACNT foams can be enhanced by introducing the mesoscale architecture without affecting the damping capacity.

\begin{figure}[t]
	\centering
	\includegraphics[width=\textwidth]{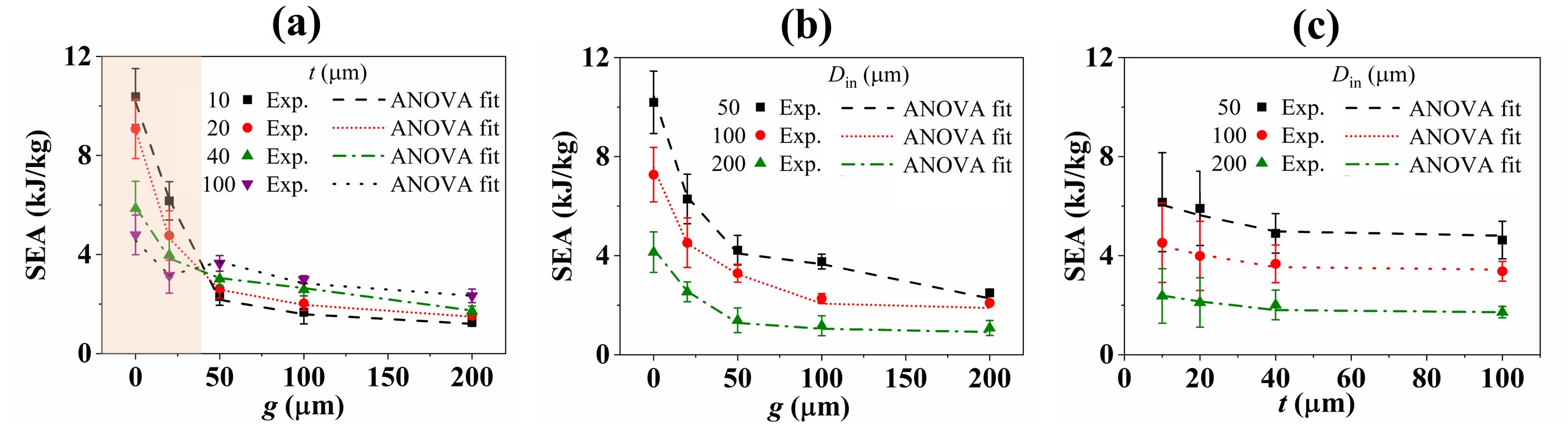} \caption{Combined effects of design variables $t:g$ (a), $D_{in}:g$ (b), and $D_{in}:t$ (c) on the response variable SEA}
	\label{her3}
\end{figure}

Following the quasistatic compression experiments, we modeled the experimental data using the Analysis of Variance (ANOVA) method to identify the most significant design variables and their interactions. From \autoref{her2}(b,c,d), it is clear that individual design variables significantly affect the quasistatic specific stress-strain response of the VACNT foam sample. It is noteworthy that the $SEA$, $\sigma_p^*$, and $E^*$ of architected VACNT foam with $D_{in}=50\;\mu m$, $t=10\; \mu m$, and $g=0$ is even higher than non-architected VACNTs. In \autoref{her3}, we plot the $SEA$ to show the effects of different combinations of design variables (in each interaction plot, $SEA$ values are averaged over the third design variable). For lower values of gap, it is noticeable from the shaded region in \autoref{her3}(a) that $SEA$ is decreasing with an increase in thickness. However, as the $g$ increases above $g > 40 \;\mu m$ , the trend reverses such that the higher $SEA$ occurs at higher thicknesses. This interaction between gap and thickness ($t:g$ interaction) for $SEA$ persists for all levels of $D_{in}$. This trend reversal is arising from competing effects between CNT morphology changes resulting from the size-confined CNT growth (depending on the $t$) and lateral interactions among adjacent cylinders (depending on both $t$ and $g$).

\begin{figure}[t]
	\centering
	\includegraphics[width=\textwidth]{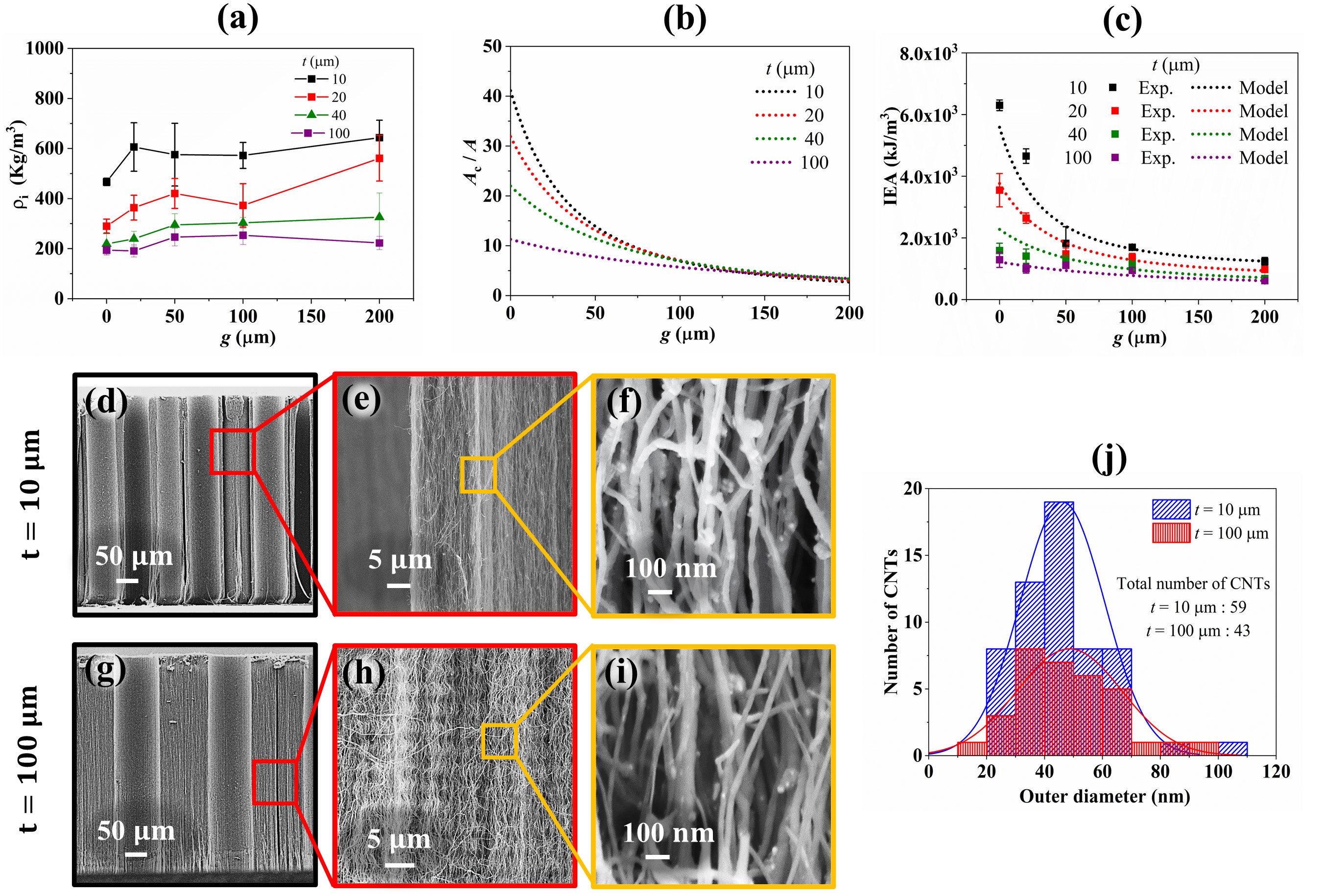} \caption{Intrinsic density of architected VACNT foams as a function of thickness ($t$) and gap ($g$) for $D_{in}=50 \mu m$. (b) Normalized total outer curved surface area of cylinders as a function of gap and thickness for $D_{in}=50\;\mu m$. (c) Intrinsic energy absorption $(IEA)$ as a function of thickness and gap for $D_{in}=50\;\mu m$. SEM images of pristine architected VACNT foams with cylinders' thickness $t=10\; \mu m$ (d,e,f) and $t=100\; \mu m$ (g,h,i) (refer \hyperref[section:sd]{SI} for SEM images of compressed samples). (j) Distribution of outer diameter of individual MWCNTs measured from SEM images.}
	\label{her4}
\end{figure}

To elucidate this intriguing effect, we calculate the intrinsic density ($\rho_i$) of architected VACNT foams by dividing the measured bulk density ($\rho_b$) by the fill factor of the cylindrical pattern $\left( \rho_i = \frac{\rho_b}{V_f}\right)$, $\rho_i$ characterizes the mass density of CNTs within the cylinder wall region i.e., excluding the volume associated with hollow and gap regions. Intrinsic density increases with decreasing thickness, signifying the emergence of a size-effect. \autoref{her4}(a) shows the effect of both gap and thickness on $\rho_i$. While the gap does not have any effects, the effect of thickness on $\rho_i$ is clearly evident. Examining the SEM images of samples for two extreme thickness values, i.e., $t=10\; \mu m$ and $t=100\;\mu m$, a clear difference in morphology is apparent (\autoref{her4}(d,g)). For $t=10\; \mu m$, the CNTs are more vertically aligned and packed more densely due to size-confined CVD growth, resulting in higher intrinsic density (\autoref{her4}(e,f)) \autocite{jeong2009effect}. For $t=100\; \mu m$, CNTs are wavier and less-dense (\autoref{her4}(h,i)). While the average diameter of individual CNTs is similar for $t=100\; \mu m$ and $t=10\; \mu m$ (\autoref{her4}(j)), the number density is much higher for $t=10\; \mu m$, resulting in higher intrinsic density. 
 
In addition to the improvement of properties from aforementioned size-effects, we further enhance the properties by exploiting increased lateral interactions among adjacent mesoscale cylinders. For a particular design thickness, dramatic increase in $SEA$ at smaller gaps is a consequence of this enhanced lateral interactions among adjacent cylinders (\autoref{her3}(a)). They also result in an overall increment of specific stress due to constrained deformation of individual cylinders. We hypothesize that the total amount of lateral interactions between cylinders must be proportional to the total cylindrical outer surface area ($A_c$). In \autoref{her4}(b), we plot $\frac{A_c}{A}$ (cylindrical outer surface area normalized by the total cross-sectional area of the sample i.e., $A=5\;mm \times 5\;mm$) as a function of gap for different values of thickness. 
To model the combined effects of intrinsic density (arising from size-effect) and lateral interactions, we derive the following expression for $SEA$ (refer to \hyperref[section:sd]{SI} for more details),

\begin{equation}\label{eq4}
SEA=\frac{1}{\rho_{i}} \delta\left[\underbrace{2 \rho_{i}}_{\textnormal {Morphology}}+\underbrace{30.29\left(\frac{A_{c}}{A}\right)^{1.125}}_{\textnormal{Lateral\; Interactions}}\;\right]
\end{equation}

Intrinsic energy absorption (IEA)---energy dissipated (in kJ) normalized by the volume of CNTs (in $m^3$) (i.e., volume of architected VACNT foam after excluding the volume of empty spaces) is given as,

\begin{equation}\label{eqI}
	IEA=SEA\times \rho_i= \delta\left[2 \rho_{i}+30.29  \left(\frac{A_{c}}{A} \right)^{1.125} \right]
\end{equation}

Where $\delta\approx 0.83\pm 0.04$ is the damping capacity (ratio of hysteretic energy dissipated in the loading-unloading cycle to the total work done on the material during loading), which we observed to be almost constant with changing architecture. Energy dissipation in VACNT foams is a property believed to be encrypted in the atomic scale frictional interactions between nanotubes \autocite{yang2011modeling,xu2010carbon,li2012viscoelasticity,suhr2007fatigue}. Thus, damping capacity does not change with mesoscale architecture. The intrinsic density $\rho_i$ is only a function of thickness, whereas $A_c$ is a function of both thickness and gap (for constant $D_{in}$). For $g\to 0$, the contribution of $\frac {A_c}{A}$ term becomes much larger than $\rho_i$, causing higher $SEA$ to occur at lesser thickness (\autoref{her4}(b)). However, for $g>100 \;\mu m$, the value of $\frac{A_c}{A}$ becomes almost constant (\autoref{her4}(b)), and the effect of $\rho_i$ dominates, causing the trend reversal of $SEA$ seen in \autoref{her3}(a). In \autoref{her4}(c), a good agreement is observed between the experimental data and our model for $IEA$ (Eq. \ref{eqI}). Also, there is no trend reversal for $IEA$, indicating the critical role of intrinsic density in making the order to switch for $SEA$.

The intrinsic density and lateral interactions also alter the effect of $D/t$ ratio---a dimensionless parameter commonly used to evaluate the load carrying capacity of hollow cylindrical structures towards crashworthy applications. $D/t$ ratio governs the deformation mechanism for compression of cylinders and consequently affects the specific energy absorption. SEA has been reported to decrease with increasing $D/t$ ratio for compression of metallic cylinders \autocite{farley1986effect}, fiber reinforced composite cylinders \autocite{feser2020effects}, and cylinders embedded in foam matrix \autocite{alia2015energy}. In our samples, when the cylinders are far apart $(g>40 \mu m)$, the SEA increases with increasing thickness (decreasing $D/t$ ratio) (\autoref{her3}(a)), which agrees with the literature. However, the SEA stays almost constant with the $D/t$ ratio when cylinders are tightly packed and undergoing lateral interactions (Figure S5, \hyperref[section:sd]{SI}). 

While the parameter interaction effects of $t:g$ is apparent from \autoref{her3}(a), the other potential parameter interactions ($D_{in}:g$, $D_{in}:t$ and $D_{in}:t:g$) are not obvious from \autoref{her3}(b) and \autoref{her3}(c). To evaluate the significance of these parameter interactions, we devise a best fit ANOVA model for $SEA$ with significance of terms adjudged by their associated $p-values$. A small $p-value$ (typically less than $0.05$) implies that the corresponding parameter interaction term significantly influences the response variable $(SEA)$ \autocite{cao2018optimize}. As parsimonious models increase the ease of interpretation, in this work we drop lower level parameter interaction terms if higher level parameter interaction terms are significant. Further, residuals from the fitted model must pass the diagnosis test of normality \autocite{royston1982extension}, homoscedasticity \autocite{breusch1979simple}, and non-correlation \autocite{durbin1971testing}. It was observed that residuals from simple ANOVA models failed to pass the diagnosis tests and thus appropriate transformations were applied over the response variable using the renowned Box-Cox transformation \autocite{box1964analysis}. The transformation applied over the response variable helps it to adhere to the assumptions of ANOVA models (Figure S7, \hyperref[section:sd]{SI}). Our ANOVA model for response variable $SEA$ is given in Eq. (\ref{eq1}).

\begin{equation}\label{eq1}
    \frac{S E A^{0.18}-1}{0.18}=-0.118+\alpha_{g_{j}: t_{k}}+\beta_{D_{i n_{i}}: g_{j}}
\end{equation}

\begin{table}[H]
\centering
\caption{ANOVA results for SEA.}
\resizebox{0.45\textwidth}{!}{%
\begin{tabular}{ccccc}
\hhline{=====}
\textbf{Source}  & \textbf{Sum Square} & \textbf{Df} & \textbf{F} & $\mathbf{p-value}$ \\ \hhline{=====} 
$\mathbf{t:g}$ & $18.998$  & $15$  & $22.462$ & $2.2\times {10}^{-16}$                               \\ \hline 
$\mathbf{D_{in}:g}$                  & $125.639$                   & $14$                             & $159.158$   & $2.2\times {10}^{-16}$                                       \\ \hline  
\end{tabular}%
}\label{tab2}
\end{table}

\begin{figure}[t]
	\centering
	\includegraphics[width=\textwidth]{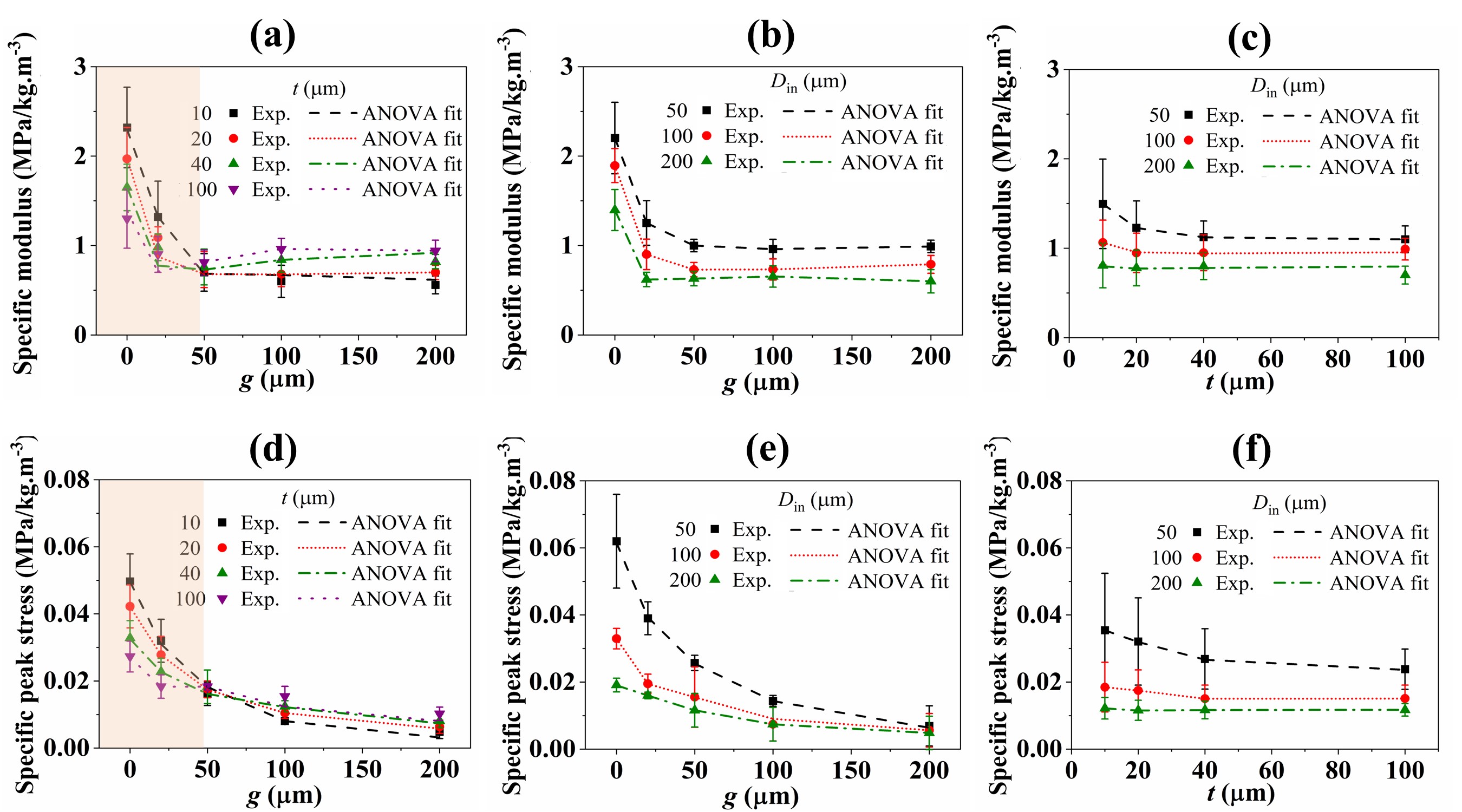} \caption{Effects of design variable interactions on response variables---(a,b,c) $E^*$ and (d,e,f) $\sigma_p^*$ }
	\label{her5}
\end{figure}

In the above ANOVA model for SEA, the parameter interaction $D_{in}:g$ also came out to be significant along with $t:g$ (Table S2, \hyperref[section:sd]{SI}). In contrast to the $t:g$, $D_{in}:g$ has no trend reversal (\autoref{her3}(b)). However, it can be noticed that at $g=0$, the SEA values are more spaced apart for different $D_{in}$ than for $g=200\; \mu m$, which could be the reason why $D_{in}:g$ parameter interaction is also significant. The $p-values$ for $t:g$ and $D_{in}:g$ are listed in \autoref{tab2}.

\autoref{her5} shows parameter interaction plots for response variables---specific peak stress ($\sigma_p^*$) and specific modulus ($E^*$). The parameter interaction between thickness and gap is very apparent and shows order switching, similar to that of $SEA$ (\autoref{her5}(a,d)). To check the significance of other parameter interactions, we use a similar strategy by formulating a simplified best fit ANOVA model with only higher-level parameter interaction terms. Analysis of residuals from simple ANOVA fits for $E^*$ and $\sigma_p^*$ also failed to pass the model diagnosis tests. This imply the need for appropriate transformations of response variables to better adhere to the assumptions of ANOVA model. In Eq. (\ref{eq2}), we show the final ANOVA model for $E^*$. The model implies that a third order parameter interaction $D_{in}:g:t$ is significant and thus all lower level parameter interaction terms are dropped. The reason for significance of third order parameter interaction could be traced to by looking at \autoref{her5}(c), which implies high variation in $E^*$ at lower levels of $t$ with respect to $D_{in}$. The best fit ANOVA model for $\sigma_p^*$ is given in Eq. (\ref{eq3}), which has a similar inference as that of $E^*$. As suggested by \autoref{her5}(f), the variation in $\sigma_p^*$ for lower levels of $t$ with each level of $D_{in}$ is also significant. Thus, the third order parameter interaction term turns out to be significant.

\begin{equation}\label{eq2}
    \frac{{E^*}^{0.626}-1}{0.626}=-0.175+\gamma_{D_{i n_{i}}:g_{j}: t_{k}}
\end{equation}
\begin{equation}\label{eq3}
    \frac{{\sigma_p^*}^{-0.141}-1}{-0.141}=1.743+\phi_{D_{i n_{i}}:g_{j}: t_{k}}
\end{equation}

In contrast, for SEA, $D_{in}$ and $t$ don't seem to interact (\autoref{her3}(c)), which is also reflected in Eq. (\ref{eq1}), where the simplified best fit ANOVA model has two second-order parameter interaction terms and cannot be further simplified to achieve the third-order parameter interaction. To investigate the absence of $D_{in}:t$ parameter interaction in SEA data, we derive the following approximate relationship among our three response variables---$SEA$, $\sigma_p^*$, and $E^*$ (refer to \hyperref[section:sd]{SI} for more details).  

\begin{equation}\label{eq7}
    SEA=\frac{\delta}{1-\delta} \times\left[{\sigma_p^*}\left[\frac{2 \epsilon_{\max }-\epsilon_{m}-\epsilon_{p}}{2}\right]-{E^*}\left[\frac{\left(\epsilon_{\max }-\epsilon_{m}\right)\left(\epsilon_{\max }-\epsilon_{p}\right)}{2}\right]\right]
\end{equation}

where, $\epsilon_{max}=0.5$, $0.48<\epsilon_m< 0.49$ (\hyperref[section:sd]{SI}), and $\epsilon_p$ is the amount of unrecovered permanent strain measured from the stress-strain response. In Eq. \ref{eq7}, the negative sign in the second term indicates that $E^*$ has a contrasting opposite effect on $SEA$ compared to $\sigma_p^*$. The absence of parameter interactions between $D_{in}$ and $t$ for SEA is likely due to cancelling of $D_{in}:t$ parameter interaction effects between $\sigma_p^*$ and $E^*$. Since the gap dramatically affects $\sigma_p^*$ more than $E^*$, the other two parameter interactions are significant for $SEA$.

\begin{figure}[t]
	\centering
	\includegraphics[width=\textwidth]{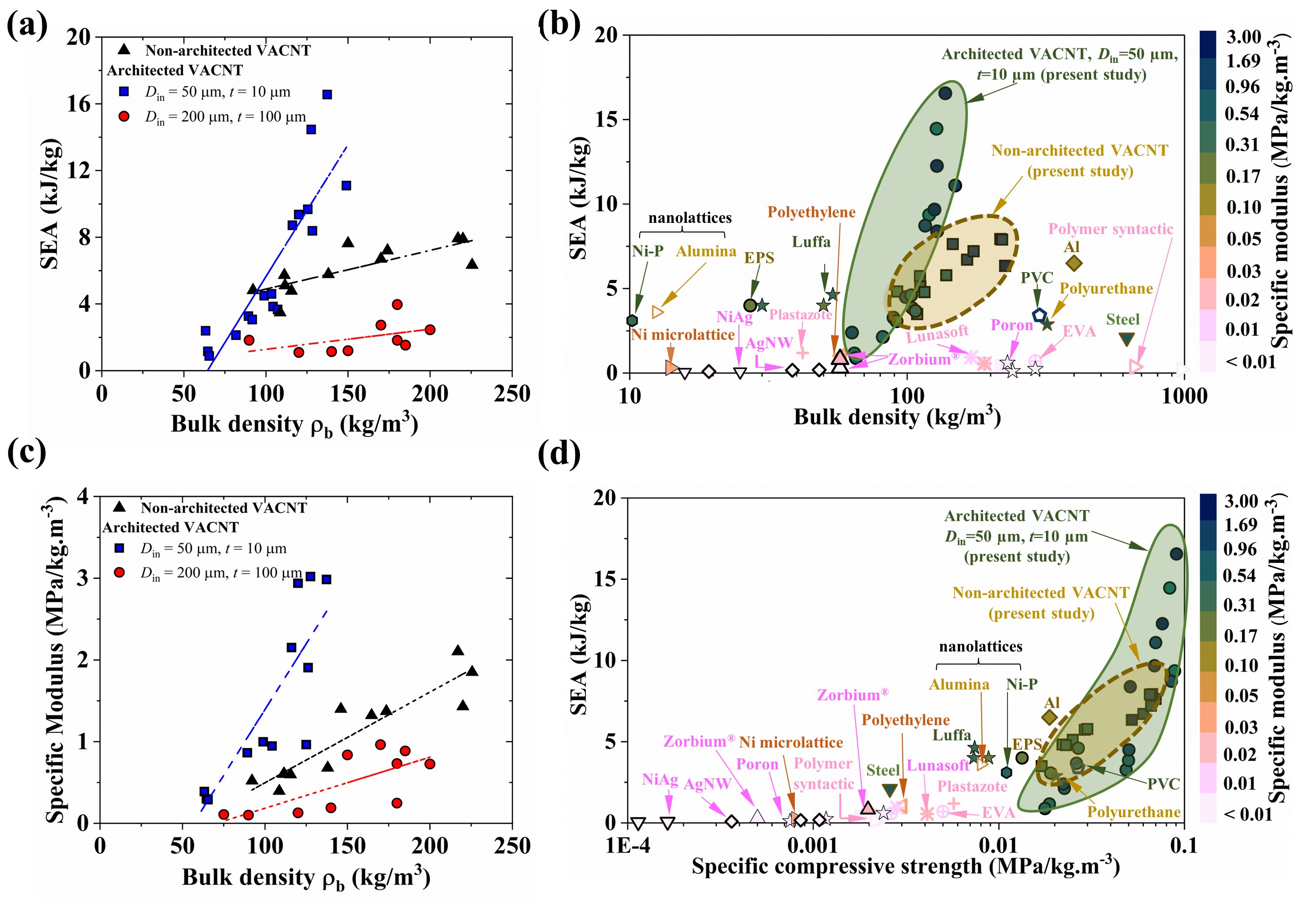} \caption{(a) Comparison of $SEA$ between architected and non-architected VACNTs; the properties are tailored by varying the gap $g$ for a given $D_{in}$ and $t$. (b) $SEA$-bulk density-specific modulus property landscape of architected and non-architected VACNT foams compared with other foams and architected materials \autocite{saha2005effect,ozturk2009energy,shariatmadari2012effects,yin2019light,jiang2015ultralight,kaya2018strain,qian2017ultralight,meza2014strong,zheng2014ultralight,castro2012synthesis,ruan2002compressive,koumlis2019strain} showing synergistic property enhancement at low density. (c) Comparison of specific modulus between architected and non-architected VACNTs. (d) $SEA$-specific compressive strength-specific modulus property landscape showing synergistic property enhancement in architected VACNTs compared to all other materials.}
	\label{her6}
\end{figure}

It is evident from parameter interaction plots and ANOVA models that we are able to achieve synergistic scaling in $SEA$, $E^*$, and $\sigma_p^*$ as functions of design variables. At the low levels of gap, all three response variables simultaneously maximize for low levels of both thickness and internal diameter. However, at the higher levels of gap, the maximization occurs for higher thickness and lesser diameters. This intriguing interplay among design variables allows drastic tailoring of mechanical properties to achieve lightweight foams for protective applications. In \autoref{her6}, we compare specific properties of architected and non-architected VACNT foams from this study with protective foam-like materials found in the literature (all properties calculated from stress-strain responses up to $50\%$ strain, if not provided directly in literature). From \autoref{her6}(a,b), it is clear that architected VACNT foams outperform non-architected VACNT foams, polymeric foams \autocite{saha2005effect,ozturk2009energy,shariatmadari2012effects}, metallic foams \autocite{yin2019light,jiang2015ultralight,kaya2018strain,qian2017ultralight,kaya2018strain}, and other architected foams \autocite{schaedler2011ultralight,meza2014strong,zheng2014ultralight} by exhibiting simultaneously improved $SEA$ and $E^*$---two often conflicting properties in materials---while being less dense. Architected VACNTs also achieve high specific compressive strength, which allows them to withstand loads without failure (\autoref{her6}(d)). In addition, architected VACNT foams have a higher range of tailorability of SEA as a function of different design variables. As seen in \autoref{her6}(b), the density and SEA are interdependent in commercial foams. Softer polymeric foams such as polyurethane \autocite{saha2005effect, koumlis2019strain}, and polyethylene \autocite{ozturk2009energy}, are less dense and have lesser $SEA$. Metallic foams such as steel \autocite{castro2012synthesis}, aluminum \autocite{ruan2002compressive}, and nickel have higher SEA but are denser. However, our architected VACNT foams surpass these limitations by exploiting structural hierarchy, size-effects from geometrically-confined CNT growth, and lateral interactions among adjacent mesoscale cylinders. For example, compared to the Zorbium (polyurethane) foams that are currently used in advanced combat helmet liners \autocite{koumlis2019strain}, our architected VACNT foams exhibit $\sim 18$ times higher SEA, $\sim 160$ times higher $E^*$, and $\sim 45$ times higher $\sigma_p^*$ at $50\%$ strain and at same quasistatic strain rate. The architected VACNT foams additionally provide exceptional thermal stability \cite{xu2010carbon,ping2019vertically} of the properties which are not possible with polymeric foams.

\section{Conclusion}

We demonstrated synergistic improvement of specific properties---compressive modulus, compressive strength, and energy absorption---by exploiting structural hierarchy, size-effects, and nanoscale inter-tube interactions in architected VACNTs. Guided by the full-factorial design of experiments (DOE) approach and the statistical analysis of variance (ANOVA) method, we found higher-order interactions among design variable of the mesoscale cylindrical architecture---leading to regimes with synergistically enhanced mechanical properties. We also showed that these intriguing parameter interactions arise from size (thickness)-dependent morphology evolutions of CNTs (number density and alignment) arising from geometrically-confined CVD synthesis and lateral interactions among adjacent cylinders tailored by the gap between them. This unique structure-property relation also alters the commonly known effects of $D/t$ ratio on thin-walled structures made of common materials and show us a novel pathway to synergistically enhance mechanical properties. Our architected VACNT foams outperform commercial polymeric, metallic, and other architected foams in terms of energy absorption, modulus, and compressive strength at ultra-lightweight. 

\section*{Acknowledgements}
This research is supported by the U. S. Office of Naval Research under PANTHER program award number N000142112044 through Dr. Timothy Bentley. We also acknowledge partial support from the Army Research Office award number W911NF2010160 through Dr. Denise Ford. The authors acknowledge the use of facilities and instrumentation at the Wisconsin Centers for Nanoscale Technology (WCNT) partially supported by the NSF through the University of Wisconsin Materials Research Science and Engineering Center (DMR-1720415). We thank all the staff of WCNT, particularly Mr. Quinn Lenord for his assistance in photolithography. 

\section*{Author contributions}

KC synthesized samples, performed compression testing, analyzed experimental results, and prepared figures. AG performed theoretical analysis while ASB provided the statistical model. RT conceived the architected-material design and supervised the project. AG and RT wrote the paper with input from all authors. All authors contributed to the interpretation of the results.

\section*{Supplementary data}

Supplementary material related to this article can be found online

\label{section:sd}

\section*{Methods}

\label{section:sd2}

First, a standard 100 mm diameter (100 crystal orientation) p-type silicon wafer was spin-coated with 10 microns thick S1813 photoresist at 3000 rpm for 30 sec and pre-baked on a hot plate at $110\;^{\circ}C$ for 45 sec to remove any solvents. After spin coating, the wafer was partially diced through the thickness (30 $\%$ of the thickness of the wafer) into $5\; mm \times 5 \;mm$ squares. Next, the diced wafer was exposed to ultra-violet (UV) light through a chrome/soda-lime photomask to transfer the micropattern. The photomask is designed with cylindrical micropatterns of various combinations of ${D}_{in}$, ${t}$, and ${g}$ and manufactured by Photo Sciences (Torrance, CA).  After 8 sec of exposure with 405 nm UV light (exposure dose of 10 $mW/cm^2$), the unexposed photoresist is removed in the 30 sec MF321 developer bath. After the developer bath, a 20 nm chromium thin film is evaporated using a metal evaporator. The remaining photoresist (exposed to UV light previously) is removed in an acetone bath, leaving a chromium film on the substrate, which prevents the growth of CNTs in the designated areas on the substrate (inverse of the architecture).

We synthesize architected VACNTs on diced patterned substrates using a floating catalyst thermal chemical vapor deposition (tCVD) process. We use a syringe pump to inject a feedstock solution of ferrocene (catalyst precursor) in toluene (carbon source) $([w/v]=0.01 \;g/ml)$ at a rate of 0.8 $ml/min$ into a furnace tube maintained at a temperature of $827\;^{\circ}C$ ($1100 K$). A mixture of argon $(95\%)$ and hydrogen $(5\%)$ flowing at 800 sccm carries toluene vapors inside the furnace, where nanotubes grow on the patterned silicon wafer. After synthesis, we remove the architected VACNT film from the furnace and cut it into squares of $5\;mm \times 5\;mm$---each square having an architecture with a specific combination of $D_{in}$, $t$, and $g$---for mechanical characterization.

	\printbibliography

@article{haghpanah2017elastic,
  title={Elastic architected materials with extreme damping capacity},
  author={Haghpanah, Babak and Shirazi, Ahmad and Salari-Sharif, Ladan and Izard, Anna Guell and Valdevit, Lorenzo},
  journal={Extreme Mechanics Letters},
  volume={17},
  pages={56--61},
  year={2017},
  publisher={Elsevier}
}

@article{fang2011revealing,
  title={Revealing extraordinary intrinsic tensile plasticity in gradient nano-grained copper},
  author={Fang, TH and Li, WL and Tao, NR and Lu, K},
  journal={Science},
  volume={331},
  number={6024},
  pages={1587--1590},
  year={2011},
  publisher={American Association for the Advancement of Science}
}

@article{yin2019impact,
  title={Impact-resistant nacre-like transparent materials},
  author={Yin, Zhen and Hannard, Florent and Barthelat, Francois},
  journal={Science},
  volume={364},
  number={6447},
  pages={1260--1263},
  year={2019},
  publisher={American Association for the Advancement of Science}
}

@article{shan2015multistable,
  title={Multistable architected materials for trapping elastic strain energy},
  author={Shan, Sicong and Kang, Sung H and Raney, Jordan R and Wang, Pai and Fang, Lichen and Candido, Francisco and Lewis, Jennifer A and Bertoldi, Katia},
  journal={Advanced Materials},
  volume={27},
  number={29},
  pages={4296--4301},
  year={2015},
  publisher={Wiley Online Library}
}

@article{lakes1996cellular,
  title={Cellular solid structures with unbounded thermal expansion},
  author={Lakes, Roderic},
  journal={Journal of materials science letters},
  volume={15},
  number={6},
  pages={475--477},
  year={1996},
  publisher={London: Chapman and Hall, c1982-2003.}
}

@article{ritchie2011conflicts,
  title={The conflicts between strength and toughness},
  author={Ritchie, Robert O},
  journal={Nature materials},
  volume={10},
  number={11},
  pages={817--822},
  year={2011},
  publisher={Nature Publishing Group}
}

@book{lakes2009viscoelastic,
  title={Viscoelastic materials},
  author={Lakes, Roderic and Lakes, Roderic S},
  year={2009},
  publisher={Cambridge university press}
}

@article{lee2014hierarchical,
  title={Hierarchical multiscale structure--property relationships of the red-bellied woodpecker (Melanerpes carolinus) beak},
  author={Lee, Nayeon and Horstemeyer, MF and Rhee, Hongjoo and Nabors, Ben and Liao, Jun and Williams, Lakiesha N},
  journal={Journal of The Royal Society Interface},
  volume={11},
  number={96},
  pages={20140274},
  year={2014},
  publisher={The Royal Society}
}

@article{meza2017reexamining,
  title={Reexamining the mechanical property space of three-dimensional lattice architectures},
  author={Meza, Lucas R and Phlipot, Gregory P and Portela, Carlos M and Maggi, Alessandro and Montemayor, Lauren C and Comella, Andre and Kochmann, Dennis M and Greer, Julia R},
  journal={Acta Materialia},
  volume={140},
  pages={424--432},
  year={2017},
  publisher={Elsevier}
}

@article{fang2021universal,
  title={Universal route for the emergence of exceptional points in PT-symmetric metamaterials with unfolding spectral symmetries},
  author={Fang, Yanghao and Kottos, Tsampikos and Thevamaran, Ramathasan},
  journal={New Journal of Physics},
  volume={23},
  number={6},
  pages={063079},
  year={2021},
  publisher={IOP Publishing}
}

@article{lin2015biomimetic,
  title={Biomimetic carbon nanotube films with gradient structure and locally tunable mechanical property},
  author={Lin, Zhiqiang and Gui, Xuchun and Zeng, Zhiping and Liang, Binghao and Chen, Wenjun and Liu, Ming and Zhu, Yuan and Cao, Anyuan and Tang, Zikang},
  journal={Advanced Functional Materials},
  volume={25},
  number={46},
  pages={7173--7179},
  year={2015},
  publisher={Wiley Online Library}
}

@article{lattanzi2014geometry,
  title={Geometry-Induced Mechanical Properties of Carbon Nanotube Foams},
  author={Lattanzi, Ludovica and De Nardo, Luigi and Raney, Jordan R and Daraio, Chiara},
  journal={Advanced Engineering Materials},
  volume={16},
  number={8},
  pages={1026--1031},
  year={2014},
  publisher={Wiley Online Library}
}

@article{copic2011fabrication,
  title={Fabrication of high-aspect-ratio polymer microstructures and hierarchical textures using carbon nanotube composite master molds},
  author={Copic, Davor and Park, Sei Jin and Tawfick, Sameh and De Volder, Michael FL and Hart, A John},
  journal={Lab on a Chip},
  volume={11},
  number={10},
  pages={1831--1837},
  year={2011},
  publisher={Royal Society of Chemistry}
}

@article{lakes1993materials,
  title={Materials with structural hierarchy},
  author={Lakes, Roderic},
  journal={Nature},
  volume={361},
  number={6412},
  pages={511--515},
  year={1993},
  publisher={Nature publishing group}
}

@article{cao2018optimize,
  title={How to optimize materials and devices via design of experiments and machine learning: Demonstration using organic photovoltaics},
  author={Cao, Bing and Adutwum, Lawrence A and Oliynyk, Anton O and Luber, Erik J and Olsen, Brian C and Mar, Arthur and Buriak, Jillian M},
  journal={ACS nano},
  volume={12},
  number={8},
  pages={7434--7444},
  year={2018},
  publisher={ACS Publications}
}

@article{durbin1971testing,
  title={Testing for serial correlation in least squares regression. III},
  author={Durbin, James and Watson, Geoffrey S},
  journal={Biometrika},
  volume={58},
  number={1},
  pages={1--19},
  year={1971},
  publisher={Oxford University Press}
}

@article{breusch1979simple,
  title={A simple test for heteroscedasticity and random coefficient variation},
  author={Breusch, Trevor S and Pagan, Adrian R},
  journal={Econometrica: Journal of the econometric society},
  pages={1287--1294},
  year={1979},
  publisher={JSTOR}
}

@article{royston1982extension,
  title={An extension of Shapiro and Wilk's W test for normality to large samples},
  author={Royston, J Patrick},
  journal={Journal of the Royal Statistical Society: Series C (Applied Statistics)},
  volume={31},
  number={2},
  pages={115--124},
  year={1982},
  publisher={Wiley Online Library}
}

@article{box1964analysis,
  title={An analysis of transformations},
  author={Box, George EP and Cox, David R},
  journal={Journal of the Royal Statistical Society: Series B (Methodological)},
  volume={26},
  number={2},
  pages={211--243},
  year={1964},
  publisher={Wiley Online Library}
}

@article{kaya2018strain,
  title={Strain hardening reduces energy absorption efficiency of austenitic stainless steel foams while porosity does not},
  author={Kaya, Ali Can and Zaslansky, Paul and Ipekoglu, Mehmet and Fleck, Claudia},
  journal={Materials \& Design},
  volume={143},
  pages={297--308},
  year={2018},
  publisher={Elsevier}
}

@article{lattanzi2015dynamic,
  title={Dynamic behavior of vertically aligned carbon nanotube foams with patterned microstructure},
  author={Lattanzi, Ludovica and Thevamaran, Ramathasan and De Nardo, Luigi and Daraio, Chiara},
  journal={Advanced Engineering Materials},
  volume={17},
  number={10},
  pages={1470--1479},
  year={2015},
  publisher={Wiley Online Library}
}

@article{saha2005effect,
  title={Effect of density, microstructure, and strain rate on compression behavior of polymeric foams},
  author={Saha, MC and Mahfuz, H and Chakravarty, UK and Uddin, M and Kabir, Md E and Jeelani, S},
  journal={Materials Science and Engineering: A},
  volume={406},
  number={1-2},
  pages={328--336},
  year={2005},
  publisher={Elsevier}
}

@article{zheng2014ultralight,
  title={Ultralight, ultrastiff mechanical metamaterials},
  author={Zheng, Xiaoyu and Lee, Howon and Weisgraber, Todd H and Shusteff, Maxim and DeOtte, Joshua and Duoss, Eric B and Kuntz, Joshua D and Biener, Monika M and Ge, Qi and Jackson, Julie A and others},
  journal={Science},
  volume={344},
  number={6190},
  pages={1373--1377},
  year={2014},
  publisher={American Association for the Advancement of Science}
}

@article{yin2019light,
  title={Light but tough bio-inherited materials: Luffa sponge based nickel-plated composites},
  author={Yin, Sha and Wang, Huitian and Li, Jiani and Ritchie, Robert O and Xu, Jun},
  journal={Journal of the mechanical behavior of biomedical materials},
  volume={94},
  pages={10--18},
  year={2019},
  publisher={Elsevier}
}

@article{meza2014strong,
  title={Strong, lightweight, and recoverable three-dimensional ceramic nanolattices},
  author={Meza, Lucas R and Das, Satyajit and Greer, Julia R},
  journal={Science},
  volume={345},
  number={6202},
  pages={1322--1326},
  year={2014},
  publisher={American Association for the Advancement of Science}
}

@article{schaedler2011ultralight,
  title={Ultralight metallic microlattices},
  author={Schaedler, Tobias A and Jacobsen, Alan J and Torrents, Anna and Sorensen, Adam E and Lian, Jie and Greer, Julia R and Valdevit, Lorenzo and Carter, Wiliam B},
  journal={Science},
  volume={334},
  number={6058},
  pages={962--965},
  year={2011},
  publisher={American Association for the Advancement of Science}
}

@article{jiang2015ultralight,
  title={Ultralight metal foams},
  author={Jiang, Bin and He, Chunnian and Zhao, Naiqin and Nash, Philip and Shi, Chunsheng and Wang, Zejun},
  journal={Scientific reports},
  volume={5},
  number={1},
  pages={1--8},
  year={2015},
  publisher={Nature Publishing Group}
}

@article{shariatmadari2012effects,
  title={Effects of temperature on the material characteristics of midsole and insole footwear foams subject to quasi-static compressive and shear force loading},
  author={Shariatmadari, Mohammad Reza and English, Russell and Rothwell, Glynn},
  journal={Materials \& Design},
  volume={37},
  pages={543--559},
  year={2012},
  publisher={Elsevier}
}

@article{ozturk2009energy,
  title={Energy absorption calculations in multiple compressive loading of polymeric foams},
  author={Ozturk, Umud Esat and Anlas, Gunay},
  journal={Materials \& Design},
  volume={30},
  number={1},
  pages={15--22},
  year={2009},
  publisher={Elsevier}
}

@article{castro2012synthesis,
  title={Synthesis of syntactic steel foam using gravity-fed infiltration},
  author={Castro, G and Nutt, SR},
  journal={Materials Science and Engineering: A},
  volume={553},
  pages={89--95},
  year={2012},
  publisher={Elsevier}
}

@article{ruan2002compressive,
  title={Compressive behaviour of aluminium foams at low and medium strain rates},
  author={Ruan, D and Lu, Guoxing and Chen, FL and Siores, Elias},
  journal={Composite structures},
  volume={57},
  number={1-4},
  pages={331--336},
  year={2002},
  publisher={Elsevier}
}

@article{qian2017ultralight,
  title={Ultralight conductive silver nanowire aerogels},
  author={Qian, Fang and Lan, Pui Ching and Freyman, Megan C and Chen, Wen and Kou, Tianyi and Olson, Tammy Y and Zhu, Cheng and Worsley, Marcus A and Duoss, Eric B and Spadaccini, Christopher M and others},
  journal={Nano letters},
  volume={17},
  number={12},
  pages={7171--7176},
  year={2017},
  publisher={ACS Publications}
}

@article{jia2019biomimetic,
  title={Biomimetic architected materials with improved dynamic performance},
  author={Jia, Zian and Yu, Yang and Hou, Shaoyu and Wang, Lifeng},
  journal={Journal of the Mechanics and Physics of Solids},
  volume={125},
  pages={178--197},
  year={2019},
  publisher={Elsevier}
}

@article{gupta2022origins,
  title={Origins of mechanical preconditioning in hierarchical nanofibrous materials},
  author={Gupta, Abhishek and Griesbach, Claire and Cai, Jizhe and Weigand, Steven and Meshot, Eric R and Thevamaran, Ramathasan},
  journal={Extreme Mechanics Letters},
  volume={50},
  pages={101576},
  year={2022},
  publisher={Elsevier}
}

@article{cao2005super,
  title={Super-compressible foamlike carbon nanotube films},
  author={Cao, Anyuan and Dickrell, Pamela L and Sawyer, W Gregory and Ghasemi-Nejhad, Mehrdad N and Ajayan, Pulickel M},
  journal={Science},
  volume={310},
  number={5752},
  pages={1307--1310},
  year={2005},
  publisher={American Association for the Advancement of Science}
}

@article{jeong2009effect,
  title={Effect of catalyst pattern geometry on the growth of vertically aligned carbon nanotube arrays},
  author={Jeong, Goo-Hwan and Olofsson, Niklas and Falk, Lena KL and Campbell, Eleanor EB},
  journal={Carbon},
  volume={47},
  number={3},
  pages={696--704},
  year={2009},
  publisher={Elsevier}
}

@article{farley1986effect,
  title={Effect of specimen geometry on the energy absorption capability of composite materials},
  author={Farley, Gary L},
  journal={Journal of Composite Materials},
  volume={20},
  number={4},
  pages={390--400},
  year={1986},
  publisher={Sage Publications Sage CA: Thousand Oaks, CA}
}

@article{alia2015energy,
  title={The energy-absorption characteristics of metal tube-reinforced polymer foams},
  author={Alia, RA and Guan, Z and Jones, N and Cantwell, WJ},
  journal={Journal of Sandwich Structures \& Materials},
  volume={17},
  number={1},
  pages={74--94},
  year={2015},
  publisher={SAGE Publications Sage UK: London, England}
}

@article{feser2020effects,
  title={Effects of transient dynamic loading on the energy absorption capability of composite bolted joints undergoing extended bearing failure},
  author={Feser, Thomas and Hassan, Jazib and Waimer, Matthias and O'Higgins, Ronan M and McCarthy, Conor T and Toso, Nathalie and Voggenreiter, Heinz and McCarthy, Michael A},
  journal={Composite Structures},
  volume={247},
  pages={112476},
  year={2020},
  publisher={Elsevier}
}

@article{thevamaran2016dynamic,
  title={Dynamic creation and evolution of gradient nanostructure in single-crystal metallic microcubes},
  author={Thevamaran, Ramathasan and Lawal, Olawale and Yazdi, Sadegh and Jeon, Seog-Jin and Lee, Jae-Hwang and Thomas, Edwin L},
  journal={Science},
  volume={354},
  number={6310},
  pages={312--316},
  year={2016},
  publisher={American Association for the Advancement of Science}
}

@article{ortiz2008bioinspired,
  title={Bioinspired structural materials},
  author={Ortiz, Christine and Boyce, Mary C},
  journal={Science},
  volume={319},
  number={5866},
  pages={1053--1054},
  year={2008},
  publisher={American Association for the Advancement of Science}
}

@article{munch2008tough,
  title={Tough, bio-inspired hybrid materials},
  author={Munch, Etienne and Launey, Maximimilan E and Alsem, Daan H and Saiz, Eduardo and Tomsia, Antoni P and Ritchie, Robert O},
  journal={Science},
  volume={322},
  number={5907},
  pages={1516--1520},
  year={2008},
  publisher={American Association for the Advancement of Science}
}

@article{tang2003nanostructured,
  title={Nanostructured artificial nacre},
  author={Tang, Zhiyong and Kotov, Nicholas A and Magonov, Sergei and Ozturk, Birol},
  journal={Nature materials},
  volume={2},
  number={6},
  pages={413--418},
  year={2003},
  publisher={Nature Publishing Group}
}

@article{chen2012biological,
  title={Biological materials: functional adaptations and bioinspired designs},
  author={Chen, Po-Yu and McKittrick, Joanna and Meyers, Marc Andr{\'e}},
  journal={Progress in Materials Science},
  volume={57},
  number={8},
  pages={1492--1704},
  year={2012},
  publisher={Elsevier}
}

@article{meza2015resilient,
  title={Resilient 3D hierarchical architected metamaterials},
  author={Meza, Lucas R and Zelhofer, Alex J and Clarke, Nigel and Mateos, Arturo J and Kochmann, Dennis M and Greer, Julia R},
  journal={Proceedings of the National Academy of Sciences},
  volume={112},
  number={37},
  pages={11502--11507},
  year={2015},
  publisher={National Acad Sciences}
}

@article{miserez2008transition,
  title={The transition from stiff to compliant materials in squid beaks},
  author={Miserez, Ali and Schneberk, Todd and Sun, Chengjun and Zok, Frank W and Waite, J Herbert},
  journal={Science},
  volume={319},
  number={5871},
  pages={1816--1819},
  year={2008},
  publisher={American Association for the Advancement of Science}
}

@article{thevamaran2015shock,
  title={Shock formation and rate effects in impacted carbon nanotube foams},
  author={Thevamaran, Ramathasan and Meshot, Eric R and Daraio, Chiara},
  journal={Carbon},
  volume={84},
  pages={390--398},
  year={2015},
  publisher={Elsevier}
}

@article{yamamoto2014thin,
  title={Thin Films with Ultra-low Thermal Expansion},
  author={Yamamoto, Namiko and Gdoutos, Eleftherios and Toda, Risaku and White, Victor and Manohara, Harish and Daraio, Chiara},
  journal={Advanced Materials},
  volume={26},
  number={19},
  pages={3076--3080},
  year={2014},
  publisher={Wiley Online Library}
}

@article{yaglioglu2012wide,
  title={Wide range control of microstructure and mechanical properties of carbon nanotube forests: a comparison between fixed and floating catalyst CVD techniques},
  author={Yaglioglu, Onnik and Cao, Anyuan and Hart, A John and Martens, Rod and Slocum, AH},
  journal={Advanced Functional Materials},
  volume={22},
  number={23},
  pages={5028--5037},
  year={2012},
  publisher={Wiley Online Library}
}

@article{raney2011tailoring,
  title={Tailoring the microstructure and mechanical properties of arrays of aligned multiwall carbon nanotubes by utilizing different hydrogen concentrations during synthesis},
  author={Raney, Jordan R and Misra, Abha and Daraio, Chiara},
  journal={Carbon},
  volume={49},
  number={11},
  pages={3631--3638},
  year={2011},
  publisher={Elsevier}
}

@article{thevamaran2015anomalous,
  title={Anomalous impact and strain responses in helical carbon nanotube foams},
  author={Thevamaran, Ramathasan and Karakaya, Mehmet and Meshot, Eric R and Fischer, Andre and Podila, Ramakrishna and Rao, Apparao M and Daraio, Chiara},
  journal={RSC Advances},
  volume={5},
  number={37},
  pages={29306--29311},
  year={2015},
  publisher={Royal Society of Chemistry}
}

@article{de2010diverse,
  title={Diverse 3D microarchitectures made by capillary forming of carbon nanotubes},
  author={De Volder, Michael and Tawfick, Sameh H and Park, Sei Jin and Copic, Davor and Zhao, Zhouzhou and Lu, Wei and Hart, A John},
  journal={Advanced materials},
  volume={22},
  number={39},
  pages={4384--4389},
  year={2010},
  publisher={Wiley Online Library}
}

@article{hutchens2010situ,
  title={In situ mechanical testing reveals periodic buckle nucleation and propagation in carbon nanotube bundles},
  author={Hutchens, Shelby B and Hall, Lee J and Greer, Julia R},
  journal={Advanced Functional Materials},
  volume={20},
  number={14},
  pages={2338--2346},
  year={2010},
  publisher={Wiley Online Library}
}

@article{fraternali2011multiscale,
  title={Multiscale mass-spring models of carbon nanotube foams},
  author={Fraternali, F and Blesgen, T and Amendola, A and Daraio, C},
  journal={Journal of the Mechanics and Physics of Solids},
  volume={59},
  number={1},
  pages={89--102},
  year={2011},
  publisher={Elsevier}
}

@article{thevamaran2014multiscale,
  title={Multiscale mass-spring model for high-rate compression of vertically aligned carbon nanotube foams},
  author={Thevamaran, Ramathasan and Fraternali, Fernando and Daraio, Chiara},
  journal={Journal of Applied Mechanics},
  volume={81},
  number={12},
  pages={121006},
  year={2014},
  publisher={American Society of Mechanical Engineers}
}

@article{needleman2012deformation,
  title={Deformation of plastically compressible hardening-softening-hardening solids},
  author={Needleman, A and Hutchens, SB and Mohan, N and Greer, JR},
  journal={Acta Mechanica Sinica},
  volume={28},
  number={4},
  pages={1115--1124},
  year={2012},
  publisher={Springer}
}

@article{liang2017compression,
  title={Compression and recovery of carbon nanotube forests described as a phase transition},
  author={Liang, Xiaojun and Shin, Jungho and Magagnosc, Daniel and Jiang, Yijie and Park, Sei Jin and Hart, A John and Turner, Kevin and Gianola, Daniel S and Purohit, Prashant K},
  journal={International Journal of Solids and Structures},
  volume={122},
  pages={196--209},
  year={2017},
  publisher={Elsevier}
}

@article{yang2011modeling,
  title={Modeling frequency-and temperature-invariant dissipative behaviors of randomly entangled carbon nanotube networks under cyclic loading},
  author={Yang, Xiaodong and He, Pengfei and Gao, Huajian},
  journal={Nano Research},
  volume={4},
  number={12},
  pages={1191--1198},
  year={2011},
  publisher={Springer}
}

@article{li2012viscoelasticity,
  title={Viscoelasticity of carbon nanotube buckypaper: zipping--unzipping mechanism and entanglement effects},
  author={Li, Ying and Kr{\"o}ger, Martin},
  journal={Soft Matter},
  volume={8},
  number={30},
  pages={7822--7830},
  year={2012},
  publisher={Royal Society of Chemistry}
}

@article{xu2010carbon,
  title={Carbon nanotubes with temperature-invariant viscoelasticity from--196 to 1000 C},
  author={Xu, Ming and Futaba, Don N and Yamada, Takeo and Yumura, Motoo and Hata, Kenji},
  journal={Science},
  volume={330},
  number={6009},
  pages={1364--1368},
  year={2010},
  publisher={American Association for the Advancement of Science}
}

@article{suhr2007fatigue,
  title={Fatigue resistance of aligned carbon nanotube arrays under cyclic compression},
  author={Suhr, Jonghwan and Victor, P and Ci, L and Sreekala, S and Zhang, X and Nalamasu, O and Ajayan, PM},
  journal={Nature nanotechnology},
  volume={2},
  number={7},
  pages={417--421},
  year={2007},
  publisher={Nature Publishing Group}
}

@article{koumlis2019strain,
  title={Strain rate dependent compressive response of open cell polyurethane foam},
  author={Koumlis, S and Lamberson, L},
  journal={Experimental Mechanics},
  volume={59},
  number={7},
  pages={1087--1103},
  year={2019},
  publisher={Springer}
}

@article{ping2019vertically,
  title={Vertically aligned carbon nanotube arrays as a thermal interface material},
  author={Ping, Linquan and Hou, Peng-Xiang and Liu, Chang and Cheng, Hui-Ming},
  journal={APL Materials},
  volume={7},
  number={2},
  pages={020902},
  year={2019},
  publisher={AIP Publishing LLC}
}
	
\renewcommand{\thefootnote}{\arabic{footnote}}

\end{document}